\RequirePackage{filecontents}      

\documentclass[11pt]{article}

\usepackage[utf8]{inputenc}   
\usepackage[T1]{fontenc}
\usepackage{lmodern}          
\usepackage{textcomp}         
\usepackage{upquote}          

\usepackage{amsmath,amssymb,stmaryrd}
\usepackage{xcolor}

\usepackage[hidelinks]{hyperref}

\usepackage{parskip}
\usepackage[square,numbers]{natbib}

\usepackage{listings}
\usepackage{newunicodechar}
\newunicodechar{λ}{$\lambda$}
\newunicodechar{ℓ}{$\ell$}
\newunicodechar{ℕ}{$\mathbb N$}
\newunicodechar{≤}{$\le$}
\newunicodechar{≥}{$\ge$}
\newunicodechar{≠}{$\neq$}
\newunicodechar{≡}{$\equiv$}
\newunicodechar{≢}{$\not\equiv$}
\newunicodechar{∩}{$\cap$}
\newunicodechar{∪}{$\cup$}
\newunicodechar{×}{$\times$}
\newunicodechar{∈}{$\in$}
\newunicodechar{∀}{$\forall$}
\newunicodechar{¬}{$\lnot$}
\newunicodechar{∨}{$\lor$}
\newunicodechar{⊥}{$\bot$}
\newunicodechar{Σ}{$\Sigma$}
\newunicodechar{→}{$\to$}
\newunicodechar{↦}{$\mapsto$}
\newunicodechar{⊎}{$\uplus$}
\newunicodechar{≟}{$\stackrel{?}{=}$}
\newunicodechar{⟦}{$\llbracket$}
\newunicodechar{⟧}{$\rrbracket$}
\newunicodechar{∷}{\texttt{::}}
\newunicodechar{₀}{\ensuremath{_{0}}}   
\newunicodechar{₁}{\ensuremath{_{1}}}   
\newunicodechar{₂}{\ensuremath{_{2}}}   
\newunicodechar{₃}{\ensuremath{_{3}}}   
\newunicodechar{₄}{\ensuremath{_{4}}}   
\newunicodechar{₅}{\ensuremath{_{5}}}   
\newunicodechar{₆}{\ensuremath{_{6}}}   
\newunicodechar{₇}{\ensuremath{_{7}}}   
\newunicodechar{₈}{\ensuremath{_{8}}}   
\newunicodechar{₉}{\ensuremath{_{9}}}   

\title{\bfseries Policy as Code, Policy as Type\\[4pt]
       \large Access Control through Dependent Typing
       \thanks {\textcopyright\ 2025 Matthew Fuchs. All rights reserved.}}
\author{Matthew Fuchs, PhD\\\texttt{mattdfuchs@gmail.com}}
\date{}

\begin{document}
\maketitle
\begin{abstract}
Policies are designed to distinguish between correct and incorrect actions; they are types. 
But badly typed actions may cause not 
compile errors, but financial and reputational harm We demonstrate how even the most complex 
ABAC policies can be expressed as types in dependently typed languages such as Agda and Lean, 
providing a single framework to express, analyze, and implement policies. 
We then go head-to-head with Rego, the popular and powerful open-source ABAC policy language. 
We show the superior safety that comes with a powerful type system and built-in proof 
assistant. In passing, we discuss various access control models, sketch how to integrate 
in a future when attributes are distributed and signed (as discussed at the W3C), and show 
how policies can be communicated using just the syntax of the language. 
Our examples are in Agda.
\end{abstract}

\section{Introduction}

With large sums, confidential data, or reputation at risk, correctly applied access policies can be the only barrier between business-as-usual and cyber disaster. The increasing complexity of the digital environment only makes this more true over time. We improve the state-of-the-art by providing provable correctness for the most complex decidable policies through the application of dependent typing. An access policy, even a complex one, is a type in a sufficiently expressive(dependent) type system, but one where the code implementing the policy can be statically type-checked. Only code obeying the policies will pass. Thus, we can provide provable guarantees a policy is correctly applied well beyond the capabilities of existing systems.

There are a number of access control paradigms available, such as access control lists (ACL)–just a triple of principal, object and right, role-based access control (RBAC)–principals have roles, roles have rights over certain objects, relation-based access control (ReBAC)–roles are just groupings and there are relationships across these groups, and attribute-based access control (ABAC)–access is a function over attributes of the accessor, the object, and the current environment, in order of increasing sophistication. Each subsumes the previous model.

We argue for the importance of ABAC for web services especially, and further that our approach represents a significant step forward beyond existing ABAC technologies. In the extended introduction we will consider 4: Rego, Sentinel, Cedar, and XACML.

Access control is implemented by technologies. These technologies require a meta-model, such as roles or relations, to describe what we want to control by modeling in our particular domain through, for example, creating roles. Consequently, what cannot be expressed in the meta-model cannot be directly protected by the technology; i.e., you cannot control what you cannot model. Where a model is too rigid to express a policy, additional, uncontrolled code must be written to surround the control technology and map into the limited model, creating a vector for bugs and attacks. If access depends on context, such as geography or time of day, or attributes of the principal or recipient, such as age, credit line, or recent withdrawals, then only ABAC is sufficient to model a policy safely.

Within ABAC solutions we can further distinguish between solutions requiring all attributes be delivered to the engine at evaluation time, meaning some other system must know how to assemble attributes from some “ground truth”, and those allowing the engine to access external sources of “truth” during evaluation, the latter decreasing the distance of policy evaluation from attribute values by one remove and simplifying the move to a distributed credential model, such as from the W3C. Solutions also divide on the expressivity of their language, between those whose policies are stated as static structures to be evaluated and those with the strength of a programming language. 

But just as you cannot protect what you cannot model, you cannot trust your model if you 
cannot be sure your code obeys it. The power of code means powerful policies can be implemented, 
but not not necessarily that they are implemented correctly. Likewise, even if individual policies are correct, 
we may still want to prove their consistent application enjoys other properties. Reduced expressive power 
can get you the first, but none of the available systems provides the second.

Needless to say, our solution, based on dependent types, provides such guarantees with the full expressive power of existing solutions. A policy evaluates a quadruple (right, accessor, object, and environment) for correctness. Alternatively stated, a policy is a \emph{type}, permitting quadruples that are correctly typed and prohibiting the ill typed. Our dependently typed approach turns this into a reality; your policies of arbitrary complexity are types defined in the language and code is statically checked to ensure only well typed (policy obeying) access will occur at run time. Furthermore, the language comes equipped with a proof assistant; not only are individual policies provable, but one can also prove invariants hold over longer stretches of code.

The Discussion section makes this argument in the abstract. If you need no convincing, you may skip to our comparison of ABAC languages. The paper then goes head-to-head with Rego, a popular open-source ABAC language/system. In it we concretely demonstrate our claims. As we include a large amount of code, this may also work as a bit of an introduction to the power of programming with dependent types, and functional programmers may choose to head straight there. We implement in Agda but everything we say holds equally in similar languages, such as Lean.

Within the body, we also demonstrate how the approach can be applied to a world of distributed truth, where evaluating a policy may involve calling external services.

\section{Discussion}

Access control systems (ACS) are the line of defense protecting your assets from the malicious and the incompetent (and the buggy). Policies are the expression of how you want that control to be applied. Policies can be expressed in many ways, from ordinary language to pseudocode to various formalisms. They can be implemented fully if your ACS supports everything expressed in your policies, or indirectly, either by dropping part of the policy, or by surrounding the ACS with additional code to fill the gap; either approach provides an attack vector for the malicious and the incompetent. Systems of increasing expressiveness have been proposed to bridge the gap between policy and implementation. 

The ideal pairing of policy language and ACS has three properties:
\begin{enumerate}
  \item \textbf{Consistency} The ACS allows no activity prohibited by a policy and prohibits no activity allowed by the policies.
  \item \textbf{Completeness} The ACS faithfully implements every expressively policy.
  \item \textbf{Sufficiency} Our policy language can express the policies we want.
\end{enumerate}

\noindent
We introduce \emph{Policy as Type} (PAT), a new approach that significantly improves the state of the art and brings us closer to that ideal state with a policy language as expressive as it is computationally feasible to obtain.

Different assets will require different policies, and there is a spectrum of meta-models, from Access Control Lists (ACLs) to Role Based (RBAC), Relationship Based (ReBAC), and Attribute Based (ABAC) technologies, in increasing order of complexity, to name the most common. Evolving needs have pushed users along this spectrum; the complexities of real-time, online, high-frequency web services and digital contracts push us to more complex policies.

The ever-increasing complexity of the digital landscape, with real-time digital contracts, distributed digital identity, digital contracts, and rampant impersonation and fraud, only makes this more difficult.

ACLs can be seen as a set of triples containing an object, a principal, and a right. It can get a bit more complicated, such as in network ACLs, where rules can apply to similar IP addresses without enumerating each of them. While a correct ACL will provide an answer to any access request, it provides no abstraction. Every triple is independent. Adding/removing/updating anything more complex than a single user or object is complex because many entries must be considered and updated. In a sense, there is neither model nor metamodel; you simply store the answers.

RBAC provides some abstraction: principals are assigned to one or more roles that they use when accessing the system. Objects, such as applications, give access rights to roles, not individuals, so one need only change an individual’s roles without affecting any objects. This significantly reduces overhead from ACLs - removing an individual means removing their roles, updating likewise means changing roles, and there is no need to consider objects. However, it cannot model simple relationships among sets of users (i.e., roles) or sets of objects.

ReBAC, which originated with Google’s Zanzibar\cite{zanzibar}, was 
developed to address this issue. Derivative systems such as 
OpenFGA\cite{openfga} and Ory\cite{ory} are based on the Zanzibar model. It 
allows easy definition of direct rights (user Alice can edit the Calendar because there is a directly defined relationship), and role based access (Bob can edit the Calendar because he’s an Editor, and Editors can edit the Calendar). However, roles can propagate through indirect relationships - Bill can edit the Calendar because he has FullAccess to the Project containing the Calendar, and there is a relationship between the Calendar and its Project. This allows for much finer control, but all these relationships are extrinsic to the users, objects, and situations it operates in - they cannot peer into the properties of objects or users.

The ABAC model adds further expressiveness by allowing the attributes of the sender, receiver, and environment to be included in a policy. Up to this point, the first two have been considered opaque - the system essentially manipulates labels applied to them - and the last has been completely absent. This allows policies to consider age, location, recent transactions, credit lines, ownership, etc.. In theory, policies of arbitrary complexity can be represented. In practice, models need to be decidable; therefore, some restrictions of power are necessary, but a policy that cannot be evaluated in finite time is itself a problem. Each ABAC system has its own compromises. We will examine four commercially available ones -  Rego, Sentinel, XACML, and Cedar - as well as PAT.

For each of these approaches there are implementing technologies. However, a technology can only control what the model can express. It may seem you can shoehorn a policy into a model, but where there is no fit, you are inevitably writing uncontrolled code to map into an inadequate model, code that leaves you open to attack. You can only control what you can model.

Our goal is to argue for ABAC as safely evaluating the most general set of policies; other models fall short of providing adequate defense. Further, we will show that even the existing ABAC languages fall short and propose a more powerful but equally implementable alternative.

Because we are particularly concerned with web services, we will consider the following kind of policy in particular: may \emph{sender} S send \emph{message} M to
\emph{receiver} R in current \emph{context} C?

More practically, we would like to be able to specify policies succinctly, evaluate them quickly, and update them easily while maintaining consistency (i.e., updating policies may create race conditions where part of a policy is implemented before a change and part is implemented after).

These practical considerations conflict with expressibility - the kinds of policies we can state in our language. Consider two:
\begin{enumerate}
    \item Children under 13 may not watch \textsf{PG-13} videos unless a
          parent grants permission.
    \item A corporate officer may transfer money from the treasury account
          \emph{only} between~09:00–17:00 US Eastern, at most five times
          per day, \emph{and} from a registered device whose request is
          signed with her key and originates inside the USA.
\end{enumerate}
  
Infringing the first may lead to legal or reputational damage, while the second may protect a digital contract against exploitation; millions of dollars have been lost through the failure to implement such constraints. How can we enforce these given our models? What can each model express?

What cannot be stated directly within some policy language, and enforced by the technology behind it, relies on the competence of programmers and attackers, meaning it is inherently unreliable. There is a dependence on the programmer, and there are additional aspects of coordination between this external code and the policy machinery. Poorly defined policies are always possible, but then it is clear where the blame lies.

\subsection{Arguing for ABAC}

With a traditional ACL, obtaining an answer is easy, but correctness is difficult to maintain. Given many users and movies, we would need to track every user’s right to view every movie. When a user turns 13, we’d need to go through all movies to grant access. Somewhere we’d need to keep a list of PG movies anyway. This little thought experiment shows ACLs have clear scaling and responsiveness issues. The problems with ACLs have given birth to our other challengers, and we will leave it here.  

With roles, we can get a bit further. Roles apply to principals not objects, but we can divide, in the first case, principals into, say, \emph{Everyone}, \emph{Over13}, and \emph{HasParentalPermission}. Every G-rated video (assuming only G and PG) allows a principal with the \emph{Everyone} role to view, while every PG-rated video only allows principals with the roles \emph{Over13} or \emph{HasParentalPermission} to view. This seems to work for a simple case, but there are still some issues. The policy language is “principals with role R can send messages of type X to object O”. Assigning principals to \emph{Over13} and \emph{HasParentalPermission} is outside the scope of the policy language; external code needs to monitor it. If permission may be given for just a single movie, then we need a \emph{PermissionToSeeMovieM} role for every movie, membership again maintained in code. If permissions can expire, we need even more code. We cannot model the (human) parent–child relationship. Each movie stands on its own; we cannot treat G movies or PG movies as groups. If something changes, we must update every affected movie individually.

Relations add still more power to the model. We can create two objects: G and PG. For any individual movie we can assign G or PG to a “rating” relationship and then inherit who can view from them. We can create a \emph{userset} of all principals and another of just those over 13. G has a “can view” relationship with both, while PG has with just the over 13 one. Each movie can now inherit the “can view” relationship from its parent, either the G or PG object. Permissions are modeled as additional permission objects with a “permission” relationship (one might normally view the relationship as proceeding from the permission object to the movie, but it is not clear from the paper whether relations can be inverted, so we will add them as relations from a given movie). The permission object would have a “permits” relation to a parent which then has a “parent” relation to the child. Zanzibar follows these using “userset rewrite rules”. We can even provide the right to view any PG movie by attaching the permission to the PG object instead of the movie.

This brings us much closer: we have a reasonable construct for designating permission, and we can connect that through parenthood with the permitted child. Expressing this in Zanzibar may be a bit tortured. However, membership in the over 13 group needs to be externally monitored, and appropriate relations added. If the permission has a time limit, that must also be externally managed. As specified, the relational policy language unions sets across relations. As such, it is difficult to describe a version of a permission that both includes parents and yet is restricted to a single child. The permission should have relations to the granting parent and the child, but also requires there be a transitive relationship - the grantor of the permission must be the parent of the child.

Considering our second proposed policy, the issue of change is even more significant; an external system needs to monitor transactions to move individuals in and out of relations, depending on temporal circumstances. These cannot be stated within the relational policy language and may change frequently. While Zanzibar has distributed consistency protocols, continually changing rights could cause significant churn.

Finally, we consider the attribute-based approach. The models we have discussed are unable to access either attributes of the entities involved (sender age, receiver rating, message signature) or the context in which a decision is being made (time of day, sender’s location). Therefore, these aspects of a policy need to be handled externally to policy technology.

We are limited to labels the system places on them for us. This requires us to create constructs to emulate the properties and execute external code to maintain them. In an ABAC model, we address them directly. We can directly access the age of a principal, we may even be able to ask about additional objects, such as if the permission property includes a permission for the given movie by another principal who is the parent of the first. 

In an ABAC system, objects and contexts have attributes we can query. Of course, the underlying system needs to model and maintain the attributes we need. But, we can just ask for the age of the sender and check if it is greater than 13. If not, we can check for, say, a permissions attribute pointing to a list of permissions, filtering to permissions pointing to the receiver (the video), and a parent attribute pointing to one or more other objects, one of which is the grantor of the permission.

In our second policy, we can consider the user’s actual credit line, how much they’ve transferred in the last 24 hours, where they are at the time of the request, etc. This lets us precisely define the constraints of our policy in the language of our implementing technology.

In theory, then, ABAC gives us access to the full power of our information system. All the information we have about sender, receiver, environment, and message, are available to us to use in evaluating a request. 

On the reasonable assumption that our underlying system can keep track of these values (the ABAC technology can only be as good as the system supplying the attributes), we then need to consider the power of the policy language under consideration.

\subsection{Comparing ABAC Languages}

We can consider four commercially available ABAC systems, as well as the one we will propose. The existing languages 
are XACML\cite{xacml}, an XML based language created 
by OASIS (there is also a YAML version called Alfa\cite{alfa}); 
Cedar\cite{cedar}, a more recent language developed by Amazon; 
Sentinel\cite{sentinel}, a language developed by Hashicorp, creators of Terraform; 
and Rego\cite{opa}, an open source 
policy language developed by Styra\cite{styra} and now part of the 
Open Policy Agent\cite{opa}.

We will compare them along the following lines:
\begin{enumerate}
    \item Where do the attributes come from?
    \item What is the expressive power of the language?
    \item How strong are its guarantees of correctness? 
\end{enumerate}

\subsubsection{Where do the attributes come from?}

The attributes a policy engine uses can either be supplied as some data structure by an outside system, such as the caller, or retrieved directly by the engine itself. In the former case, the results achieved from the policy engine can only be as good as the external system assembling the input, while in the latter the attributes are as good as the underlying system itself. While we can try to drive out sources of doubt, attribute values need to come from somewhere, some ground truth that we need accept as correct.

Languages in the former include XACML, Cedar, and Rego. A call to any of these includes a data structure, XML or JSON, with the necessary attributes, so the policies are actually a function over this data structure. However the creation of this data structure is problematic: it is beyond the control and verification of the policy apparatus; it relies on ensuring the attribute generation process (or processes) evolves correctly with the policies; and it is vulnerable to the time of check vs time of use race condition (TOCTOU). Between the time the attributes are gathered and when the policy is evaluated, values may have changed. Collecting attributes during evaluation reduces the time delay, but the situation is really akin to atomic transactions in databases. 

Sentinel has the ability to make http calls during the evaluation of a policy, meaning it can go directly to the sources of attribute values. As such it is not dependent on a data structure of dubious quality and policies are free to use any attributes supplied by an http call. However there is no control on the authenticity of these URLs or the structure of the data returned. 

PAT, being implemented in a strong, dependently typed language, can access any external data to locate attributes, like Sentinel, but the safety of those endpoints is explicitly managed. In Agda, for example, an external call whose results are just accepted as true is a \emph{postulate}, distinguishing what is true because it has been proven, and what we will assume is true (i.e., external data). With the full power of monastic functional programming we can look at deeper solutions to issues like TOCTOU. This is a problem which will only be exacerbated in a more distributed world.

\subsubsection{What is the expressive power of the language}

XACML and Cedar express policies as static structures containing some relationship information and some logic over attribute values. As such the language expressed directly in these two systems is quite limited. Any additional computation must be pushed back to the calling system.

Rego and Sentinel, on the other hand, both derive from datalog, a descendant of Prolog. Policies are arranged in logic-based rules capable of calling each other. Each rule may have several bodies, with the system using backtracking across multiple rules with multiple bodies to find a final answer.

This comes close to the power of a general purpose programming language, but these languages are deliberately weakened to ensure they terminate with an answer for each call. Sentinel’s ability to make http calls can certainly create trouble here, as backtracking could lead to multiple calls with potentially different answers each time.

PAT fits into the same arena. The underlying language - currently Agda, but eventually Lean - is deliberately designed to terminate. Indeed, all recursive calls and loops are analyzed to ensure they terminate in a finite number of steps. It is possible to program with potentially infinite data structures, but that then becomes a clear choice.  However, dependently typed languages do not have backtracking as a resolution method.

This does not seem to cause any issues, as no one wishes to create policies which cannot be evaluated in finite time.

Being directly tied to an existing language with all the power that entails, PAT has access to the full power of the underlying language as well as developments in other fields (for example, extensive areas of mathematics have been formalized in both Agda and Lean), while Rego and Sentinel are restrained in the libraries and other code they can import.

\subsubsection{How strong are its guarantees of correctness?}

Among our four existing languages, Cedar has gone the farthest in attempting to prove the correctness of its implementation, having started with a provably correct implementation and then carefully rewriting that in Rust. While XACML predates more recent efforts in provable correctness, it could certainly be subjected to a similar methodology. In both cases we should be able to assume they will correctly apply a policy as it is written.

Rego and Sentinel both trade off correctness for power. They adhere to the concept of \emph{policy as code}, directly inspired by the idea of \emph{infrastructure as code}. While they are more expressive than Cedar and XACML, policies expressed in Rego and Sentinel are simply programs, like any other programs (although based on a logical framework with backtracking unfamiliar to most programmers). Any sufficiently complex policy will be derived elsewhere and translated into Rego.  They are as vulnerable to bugs as any other program with only testing and the strength of their type systems to save the developer, and neither has a particularly strong type system.

PAT, on the other hand, exploits powerful advances in programming languages based on dependent types. In PAT, a policy is a \emph{type}, a very sophisticated type, but one for which there are type checkers. Your policies are not only encoded in the very expressive logical language of Agda’s types, but you are required to provide \emph{constructors to define} how to prove a term is a member of that type (playing a role similar to rules in Rego and Sentinel). And then you will be required to prove your code follows the policies using these constructors. The policies specify the set of legal quads (sender, receiver, message, context) and the built-in type-checker/proof assistant will ensure code actually implements the policies. PAT cannot entirely eliminate programmer error, but way fewer can pass through..

Beyond this, Agda and Lean both double as proof assistants. Not only can one , for example, prove your policies are consistent and that your code obeys them individually, one can prove that larger chunks of code interacting with multiple policies will still maintain invariants. One could, for example, encode ReBAC constraints into your policies and then directly interact with Zanzibar or a clone by postulating they can correctly provide the relation between two objects. Postulates are a way to work with foreign functions (since, being foreign, they cannot be checked with Agda itself).

\subsection{From here}

From here we will explain our approach in a head to head comparison with Rego, currently the open source champion of the ABAC approach. This will involve a lot of Agda code and can also be seen as a bit of an introduction to the value of dependently typed programming. We start with a translation of the introductory example from the Rego documentation and move on from there. We implement the movie policy in detail and describe how this can work in a decentralized trust environment, such as envisioned by the W3C Verifiable Claims WG.

\section{Dependent types and propositions as types - turning a policy into a type}

There is a new family of programming languages called dependently typed.
This family several languages in active use, including Agda\cite{plfa}, Lean\cite{lean4book}, 
and Idris\cite{idris2tutorial}, among others.
Our examples will use Agda, but our reasoning applies to all members of the family.  Because dependent types are unfamiliar to most programmers,
 we will spend some time describing a particular class of dependent types called propositional types which extend the familiar structural types, 
 and then use those to rewrite the Rego introductory example in Agda.

Dependent typing breaks down the barrier between types and terms deeply rooted in 
other programming languages. The canonical example is vectors of length \emph{n}, 
defined as:

\begin{lstlisting}
data Vec  : Set a : ℕ → Set a where  
  []  : Vec A zero  
  _::_ : ∀ x : A (xs : Vec A n) → Vec A (suc n)
\end{lstlisting}

Here \texttt{A} gives the type of entries in the \texttt{Vector}. 
Bypassing much subtlety, \texttt{Set} is Agda-speak for a type, so given a type \texttt{A}, \texttt{Vec A} is a type constructor which, 
given a natural number, \emph{n}, creates the type \texttt{Vec A }\emph{n}. 
(ℕ is the set recursively consisting of \texttt{zero} and \texttt{(suc n)}, where \emph{n} is a member of ℕ.)  
There are two constructors, \texttt{[]} creates the unique term of type 
\emph{Vec A 0}, and \texttt{\_::\_} adds an object of type \texttt{A} to the front of a 
vector of type \texttt{Vec A }\emph{n}, creating a vector of size \emph{n}\texttt{ + 1}, of 
type \texttt{Vec A (suc }\emph{n}\texttt{)}. This shows a dependent type, a type containing a term, in this 
case an ℕ, so \texttt{‘a’ :: ‘b’ :: []} is of type \texttt{Vec Char 2}. In most languages 
programmers are used to, you cannot have a value, such as 2, as part of a type definition. 
Note that the constructors have complete control over \emph{n} - you can only declare 
an object of \texttt{Vec A }\emph{n} by constructing it one object at a time.

(Note that \texttt{Set} is at the bottom of a hierarchy of types so the type system avoids Russell’s paradox. This is not germane to the current work, but the curious reader is certainly encouraged to find out more.)

More interesting are types such as 
\begin{lstlisting}
data _≤_ : Rel ℕ 0ℓ where  
  z≤n : ∀ {n}                 → zero  ≤ n  
  s≤s : ∀ {m n} (m≤n : m ≤ n) → suc m ≤ suc n
\end{lstlisting}

\texttt{\_≤\_} is a relation between two natural numbers. This describes types such as \texttt{5 ≤ 6}. For someone just exposed to this approach that takes a while to sink in.  \texttt{5 ≤ 6} is a type, not a function from a pair of numbers to a boolean value. How do we get such a thing? We have two type constructors, \texttt{z≤n} and \texttt{s≤s}. The first takes a single number, \emph{n}, and constructs the type \texttt{zero ≤ n} for whatever \emph{n} is. The curly braces indicate that Agda will attempt to infer the appropriate \emph{n}, so you don’t need to do so. Clearly that won’t give us an example of \texttt{5 ≤ 6}, so we look to the other constructor. It says, if we have an object of type \texttt{m ≤ n} then we can get back an object of type \texttt{suc m ≤ suc n}. In this case \texttt{m} is 4 and \texttt{n} is 5. We need (\texttt{s≤s} …\emph{object of type 4} ≤ \emph{5}). How do we get an object of type \texttt{4 ≤ 5}? We repeat, until we can get down to using \texttt{z≤n}, at which point \texttt{n} is just \texttt{1} (i.e., \texttt{6 - 5}). The (only) object of type \texttt{5 ≤ 6} (given these definitions) is \texttt{(s≤s (s≤s (s≤s (s≤s (s≤s z≤n {1}))))}). In general Agda infers the argument to \texttt{z≤n}, it is an \emph{implicit} argument.  We supply it here for clarity; the curly braces indicate to the compiler we are replacing an implicit parameter. The object is a proof that \texttt{5 ≤ 6}.

Another, almost ubiquitous, example is the equivalence type constructor (\_≡\_), defined as   
\begin{lstlisting}
data _≡_ {A : Set} (x] : A) : A → Set where  
  refl : x ≡ x  
\end{lstlisting}
Equivalence takes an object, \emph{x}, of (implied) type \texttt{A} and constructs the type \texttt{x ≡ x}. At first glance this might appear trivial, but it is quite subtle. We can write the type \texttt{x ≡ y}, and try to construct an object of that type with \texttt{refl}, but it will only type check if we can convince Agda that \emph{x} and \emph{y} are always the same. It turns out needing to show two objects with different definitions are actually the same happens a lot, and we will see examples shortly. (Because dependently typed languages allow terms into type definitions, they also allow function calls and other constructs. These may need to be executed while type checking, sometimes leading to long compile times. However Agda, and similar languages, forces iteration and recursion to terminate so compile times are finite. Agda has other constructs to support controlled non-termination.)

We will dive deeper into this as we look at defining policies through dependent types. The introductory example in the Rego documentation, which we rewrite in Agda, is a filter on a putative server configuration. The configuration is expressed in YAML and the Rego code walks the tree.  As given, the typing of structures and \texttt{Server}, \texttt{Protocol}, \texttt{Port} and \texttt{Network} is entirely implicit in the Rego code, although that can also be validated using JSON-Schema.

We now duplicate the Rego example with the following Agda code. Agda, like Haskell and many other languages (both functional and non-functional), is strongly typed, so type declarations missing from Rego are present. 
\begin{lstlisting}
data Protocol : Set where  
   https : Protocol  
   ssh : Protocol  
   mysql : Protocol  
   memcache : Protocol  
   http : Protocol  
   telnet : Protocol

data Network : Set where  
   network : String → Bool → Network

data Port : Set where  
   port : String → Network → Port

data Server : Set where  
   server : String → List Protocol → List Port → Server

protocols : Server → List Protocol  
protocols (server _ a _) = a

net1 = network "net1" Bool.false  
net2 = network "net2" Bool.false  
net3 = network "net3" Bool.true  
net4 = network "net4" Bool.true

publicNetwork : Network → Bool  
publicNetwork (network _ a) = a

getNetwork : Port → Network  
getNetwork (port _ n) = n

p1 = port "p1" net1  
p2 = port "p2" net3  
p3 = port "p3" net2

app = server "app" (https ∷ ssh ∷ []) (p1 ∷ p2 ∷ p3 ∷ [])  
db = server "db" (mysql ∷ []) (p3 ∷ [])  
cache = server "cache" (memcache ∷ []) (p3 ∷ [])  
ci = server "ci" (http ∷ []) (p1 ∷ p2 ∷ [])  
busybox = server "busybox" (telnet ∷ []) (p1 ∷ [])

servers = app ∷ db ∷ cache ∷ ci ∷ busybox ∷ []  
networks = net1 ∷ net2 ∷ net3 ∷ net4 ∷ []  
ports = p1 ∷ p2 ∷ p3 ∷ []
\end{lstlisting}
We first define the data types and then define the values from the example. 

The Rego example then introduces two policies:

1. Servers reachable from the Internet must not expose the insecure 'http' protocol.  
2. Servers are not allowed to expose the 'telnet' protocol.

We will slightly generalize this and consider 3 lists of protocols:

1. \texttt{badProtocols}, which just includes \texttt{telnet}
\begin{lstlisting}
   badProtocols : List Protocol  
   badProtocols = telnet ∷ []  
\end{lstlisting}   
2. \texttt{weakProtocols}, which just includes \texttt{http}
\begin{lstlisting}  
   weakProtocols : List Protocol  
   weakProtocols = http ∷ []  
\end{lstlisting}
3. \texttt{strongProtocols}, which contains everything else  
\begin{lstlisting}
   strongProtocols : List Protocol  
   strongProtocols = https ∷ ssh ∷ mysql ∷ 
        memcache ∷ http ∷ []
\end{lstlisting}

So far Agda looks like any conventional functional language, like Haskell. We now wade into the weeds.

We have a bunch of servers and we can define three properties of servers:

1. A \texttt{Server} is good if it exposes no bad protocols  
2. A \texttt{Server} is private if it is not exposed to the Internet  
3. A \texttt{Server} is compliant if it is good and either it is private or it exposes no weak protocols

They appear in Agda code as:
\begin{lstlisting}
data GoodProtos : Server → Set where  
    *good* : (s : Server) → 
             ((protocols s) ∩ badProtocols) ≡ [] → GoodProtos s

data PrivateServer : Server → Set where  
    *private* : (s : Server) →   
(Data.List.Base.length (anyExposed (portList s))) ≡ 0 
             → PrivateServer s

data GoodServer : (s : Server) → Set where  
   *goodserver* : ∀ (s : Server) → (GoodProtos s) →   
 	(PrivateServer s) → (GoodServer s)  
   *safeserver* : ∀ (s : Server) → (GoodProtos s) →   
                  ¬ (PrivateServer s) →   
                 ((protocols s) ∩ weakProtocols ≡ []) →   
                 (GoodServer s)
\end{lstlisting}
We now start using dependent types in earnest, and they show their power. Here, the types serve as propositions about \texttt{Servers}.  \texttt{GoodProtos s} represents that we have a \texttt{Server} none of whose protocols are prohibited (\texttt{protocols} is a helper function).  It is dependent because the type \emph{depends} on a term, in this case the particular \texttt{Server}, \texttt{s}. Since it is dependent on the particular \texttt{Server}, \texttt{GoodProtos db} is not the same type as \texttt{GoodProtos app}.

(The helper function protocols is defined as  
\begin{lstlisting}
protocols : Server → List Protocol  
protocols (server _ a _) = a  
\end{lstlisting}
and just returns the list of \texttt{Protocol} terms defined for the server.)

To construct a value of type \texttt{GoodProtos}, we need to call a constructor.  There is only one, \texttt{*good*}. We need to pass it first a \texttt{Server}, \texttt{s}, and then something of \emph{type} \texttt{(protocols s) ∩ badProtocols) ≡ []}, at which point we receive back a value of type \texttt{GoodProtos}. Look at the definition of the \texttt{Server} name \texttt{db}. We can get a proof of \texttt{GoodProtos db} with the statement \texttt{*good* db refl}. Since the definition of \texttt{db} is known at compile time, Agda can automatically reduce \texttt{((protocols s) ∩ badProtocols)} to \texttt{((mysql ∷ []) ∩ telnet ∷ [])} and then to \texttt{[]}, at which point \texttt{refl}, implicitly given the result of this operation (the Agda compiler can infer many arguments automatically), generates a value of type \texttt{[] ≡ []}. This type checks, and we have  proof that \texttt{db} does not expose any bad protocols.  You can enter \texttt{*good* db refl} into the interpreter or compile it and it type checks; it is a value of type \texttt{GoodProtos db}.  However, you cannot do the same with \texttt{busybox}, which exposes \texttt{telnet}.  Because the intersection is no longer empty, you would need to create a value of type \texttt{(telnet ∷ []) ≡ []}. But that type is empty - the only constructor for equivalence is \texttt{refl}, which only takes one value.  You cannot create a value of type \texttt{GoodProtos busybox}, so \texttt{*good* busybox refl} does not type check.

Getting a value of \texttt{PrivateServer s} proceeds similarly, although here we ensure the number of exposed ports is 0.  Once again, a value of the appropriate type can only be established if, indeed, there are no ports on the server exposed to the Internet.

Finally, a \texttt{Server}, \emph{s}, is acceptable if we have a proof of \texttt{GoodServer s}.  There are two possibilities here: only acceptable protocols are used and either the server is \texttt{Private} or it is not \texttt{Private} but no weak protocols are exposed.

While it is good to prove such properties, an object of type \texttt{GoodServer s} is fairly opaque and doesn’t give access to \emph{s} at the term level. We really want both the object, \emph{s}, and the proof together.

A dependent pair is the combination of an object and a statement about that object.  The type   
\texttt{Σ[ s ∈ (Server) ](GoodServer s)} is an existentially quantified type. It says here exists \emph{s}, a \texttt{Server}
 which is a \texttt{GoodServer}. The constructive logic underlying the type system requires me to provide both such an object and a proof. The constructor for a dependent pair is \texttt{\_,\_}.  An example is:
\begin{lstlisting}
(db , *goodserver* db (*good* db refl) (*private* db refl)
\end{lstlisting}
where \texttt{db} is defined as above. This type checks, meaning I have a pair of \texttt{db} and a proof that \texttt{GoodServer db} holds.

From this, one can prove servers are good at construction time.  For example, in 
\begin{lstlisting}
randomServer : Σ[ s ∈ (Server) ] (GoodServer s)  
randomServer = let foo = server "foo" (mysql ∷ []) 
                                (p3 ∷ []) in   
     (foo , *goodserver* foo (*good* foo refl) 
               (*private* foo refl))
\end{lstlisting}
\texttt{randomServer} is a dependent pair. One can get the \texttt{Server} object by projection:  
\begin{lstlisting}
goodServer = proj₁ randomServer  
\end{lstlisting}
and be guaranteed to have a usable \texttt{Server}. 

This immediately contrasts with Rego, where the decision that a server is good lies far from its definition. With Rego, the definition of a \texttt{Server} must be serialized  into JSON or other understood format, transmitted to a Rego server with a query, the answer received and interpreted, and finally some decision must be made. You cannot bring the policy into the construction of the \texttt{Server}, only through a convoluted check.

As argued above, if you create your \texttt{Server} where the compiler can “see” it, it can move policy checking into a compile time action. However, it may be the \texttt{Server} configuration is loaded or in a separate module. Then the compiler cannot see the configuration and we need to deal with a list of \texttt{Servers} which may be either good or not. The function \texttt{goodServerCheck} examines a \texttt{Server} and returns a \texttt{Maybe (Σ[ s ∈ (Server) ] (GoodServer s))} - either a dependent pair or \texttt{nothing}. 
\begin{lstlisting}
goodServerCheck : (s : Server) → 
                   Maybe (Σ[ s ∈ (Server) ] (GoodServer s))  
goodServerCheck s@(server _ protList portList) with   
     (emptyIntersection protList badProtocols)  
...       | no _ = nothing  
...       | yes noBad with isPrivate s  
...              | yes yesPrivate = just (s , *goodserver* s   
                                 (*good* s noBad) yesPrivate)  
...              | no notPrivate with (emptyIntersection protList            
                                                     weakProtocols)  
...                      | no _ = nothing  
...                      | yes noWeak = just (s , *safeserver* s   
                                (*good* s noBad) notPrivate noWeak)
\end{lstlisting}
Here, \texttt{emptyIntersection} is \texttt{Dec([] ≡ [])}. If it fails on the 3rd line, then return \texttt{nothing}. Otherwise, check if it’s private. If so, return a pair, using \texttt{*goodserver*} for proof.  The values \texttt{noBad} and \texttt{yesPrivate} are evidence for the proof.  If it’s not private, then test for \texttt{weakProtocols}. If there are none, return a pair, using \texttt{*safeserver*} as proof - or \texttt{nothing}.

With this function we can get either the list of good \texttt{Server}s or of violations by simply filtering the list. For the good servers one might return the list with a proof that all servers in the list are good. 
\begin{lstlisting}
goodServerList : List Server → List Server  
goodServerList [] = []  
goodServerList (h ∷ t) with goodServerCheck h  
...                      | nothing = goodServerList t  
...                      | just _  = h ∷ (goodServerList t)
\end{lstlisting}
There are a few additional desiderata to consider. 

As mentioned, \texttt{emptyIntersection} is \texttt{Dec([] ≡ []}). Decidable, \texttt{Dec}, is defined as
\begin{lstlisting}
data Dec {a} (P : Set a) : Set a where  
  yes : ( p :   P) → Dec P  
  no  : (¬p : ¬ P) → Dec P
\end{lstlisting}
\texttt{Dec} takes a propositional type and has two constructors.  \texttt{Yes} returns with a proof of type \emph{P} and \texttt{no} returns with a proof of its negation. For a propositional type to be decidable then one must be able to prove it holds or prove it doesn’t. 

There are two ways to prove a \texttt{Server} is a \texttt{GoodServer}. One, \texttt{*safeserver*}, requires proving a \texttt{Server} is  \texttt{¬(PrivateServer s)}. This requires a brief detour into the treatment of negation in Agda. 

Proof in Agda is based on the theories of Per Martin-Löf (for a short presentation consider this 
preprint\cite{rijke-hott}).  
As it is a constructive logic, it lacks the law of the excluded 
middle, i.e., \texttt{(P ∨ ¬P)} is not an axiom. Instead, \texttt{¬(PrivateServer s)} is shorthand for  \texttt{(PrivateServer s) → ⊥}.  
\texttt{⊥} (bottom) is the empty type, and deriving it means there’s been a contradiction, i.e., from bottom you can prove anything. 
Essentially, \texttt{¬(PrivateServer s)} means you should not be able to prove \texttt{(PrivateServer s)}, 
because if you could, that would be a contradiction.

While Agda does not employ the law of the excluded middle for proof, there still are types that are decidable, 
as just mentioned. For example, a natural number is either 0 or it isn’t, so there is a 
decidable version of equivalence for ℕ: \_≟\_. A value of this type is either a \texttt{yes} or a \texttt{no}, and we use that in the definition of \texttt{isPrivate}, where we check how many exposed ports the server has:  
\begin{lstlisting}
isPrivate : (s : Server) → Dec (PrivateServer s)  
isPrivate s with (Data.List.Base.length 
                      (anyExposed (portList s))) ≟ 0  
...          | yes eq = yes (*private* s eq)  
...          | no neq = no (negatePrivateServer s neg)  
\end{lstlisting}
Being decidable, \texttt{(Data.List.Base.length (anyExposed (portList s))) ≟ 0} actually returns either \texttt{yes 0 ≡ 0} or \texttt{no (0 ≡ 0 → ⊥)}. We will use both.

The first case uses \texttt{(*private* s eq)} to return a value of \texttt{PrivateServer s}. For \texttt{¬(PrivateServer s)} we consider the helper function \texttt{negatePrivateServer}, defined as
\begin{lstlisting}
negatePrivateServer : ∀ (s : Server) →   
     {Data.List.Base.length (anyExposed (portList s)) ≢ 0 } →   
     ¬ (PrivateServer s)  
negatePrivateServer s { neq } (*private* _ p) = ⊥-elim (neq p)
\end{lstlisting}

As explained, \texttt{¬(PrivateServer s)} is shorthand for \texttt{(PrivateServer s) → ⊥} 
and \texttt{(Data.List.Base.length (anyExposed (portList s)) ≢ 0} is really 
shorthand for \texttt{0 ≡ 0 → ⊥}. In the body of \texttt{negatePrivateServer} we 
first pass in a function that derives a contradiction when handed \texttt{0 ≡ 0} 
(we have this from \texttt{neg} in \texttt{isPrivate}, as described above).  That 
is then curried to a function of type \texttt{Private p → ⊥}. 
\texttt{(*private* \_ p)} is the parameter to that function (because, being 
\texttt{*private*}, \texttt{p} cannot contain any exposed ports) and 
\texttt{⊥-elim (neq p)} derives the contradiction we need.

\subsection{Regression Tests vs. Regression Proofs}

Programmers are all familiar with the concept of regression tests - a set of tests run after changes to try to ensure that new code doesn’t break existing behavior. Regression tests basically run sets of sample data on various parts of the application to try to ensure behavior is correct.  Likewise we can prove properties of our current system which may fail if it is changed.

Just as we can prove properties of a transaction request, we can prove properties of the underlying model. For example, we have established a \texttt{Protocol} type with a number of different protocols. We have divided them into three categories, \texttt{badProtocols}, \texttt{weakProtocols}, and \texttt{strongProtocols}.  We might want to ensure every \texttt{Protocol} is in one of these three categories, and that they are exclusive. We can prove these properties for the current code, and if a \texttt{Protocol} is added but not categorized, or if a \texttt{Protocol} is entered in two categories, the code proving these properties will not compile. This is a regression \emph{proof}, not just a test.

To prove the three categories are exclusive do as follows:  
\begin{lstlisting}
separate : (badProtocols ∩ weakProtocols ≡ []) × 
                  (badProtocols ∩ strongProtocols ≡ [])  
                × (weakProtocols ∩ strongProtocols ≡ [])  
separate = refl , refl , refl  
\end{lstlisting}
The type of \texttt{separate} is an \emph{and} over the three conditions.  This is represented by × at the type level.  We generate an object of the type by proving the type, which we do with three repetitions of \texttt{refl}.  For each of these, the compilation has evaluated the condition (e.g., \texttt{badProtocols ∩ weakProtocols} and passed the results to the only constructor for equivalence, \texttt{refl}. Since the intersections are all empty, this proves the proposition.

Proving each \texttt{Protocol} is in one category or another requires a proof for each:  
\begin{lstlisting}
allProtosAssigned : (p : Protocol) → 
        ((p ProtDecSetoid.∈ badProtocols) ⊎   
         (p ProtDecSetoid.∈ weakProtocols) ⊎   
         (p ProtDecSetoid.∈ strongProtocols))  
allProtosAssigned https = inj₂ (inj₂ (here refl))  
allProtosAssigned ssh = inj₂ (inj₂ (there (here refl)))  
allProtosAssigned mysql = inj₂ (inj₂ (there (there 
                                        (here refl))))  
allProtosAssigned memcache = inj₂ (inj₂ (there (there 
                                  (there (here refl)))))  
allProtosAssigned http = inj₂ (inj₁ (here refl))  
allProtosAssigned telnet = inj₁ (here refl)  
\end{lstlisting}
Here, the symbol ⊎ represents an \texttt{or}, so a disjoint union type.  To follow the first case, for \texttt{https}, we first need to prove it is in \texttt{strongProtocols}.  After Agda is done applying implicits, the calls to  \texttt{inj₂ (inj₂)} indicate we are looking at the second type of the second union, i.e., \texttt{strongProtocols}. The call \texttt{here refl} grabs the first item in the list and proves it’s equivalent to \texttt{https}. For the next entry, \texttt{there} grabs the tail of the list, and so on.

Were another protocol to be added to the type, say \texttt{smtp}, and not included in one of our lists, or included in more than one, these proofs would fail, just as a regression test may fail.

These proofs are small but ensure an important property. More complex proofs can be applied to other parts of the system. When the script is monitoring actions in a financial contract, proving certain conditions hold (or do not) can be worth millions of dollars.

\section{Approving a transaction}

Generically, a transaction, such as a call from an entity to a service or a credit card purchase, involves multiple parties.  Each party has some set of attributes (we may also call these properties or, to foreshadow, claims). A policy needs to determine if the transaction is valid or not. To do so it evaluates some proposition for each of the parties, and if all are proven, then the transaction can proceed. The value of dependent types is we can logically specify the policies in code and enforce them in code.

Consider a service providing videos with a simple policy: a user must be older than a video’s minimum age to see it, or have permission from their parent to do so. Just to add a little networking, we require that the request pass over https before noon and the answering server is  approved. Finally, we want to establish that the returning video is the one requested.

We define Agda record types for the various parties to the transaction. Each record type has a field for a unique name and additional fields to provide attributes. For our purposes, we assume we a reference to atrusted external system to provide the appropriate attributes given a unique name. 
\begin{lstlisting}
record ItemRequest : Set where  
 field  
   name : String  
   videoName : String  
   ageLimit : ℕ

record Item : Set where  
 field  
   name : String  
   ageLimit : ℕ

{-# NO_POSITIVITY_CHECK #-}  
record User : Set where  
 inductive  
 field  
   name : String  
   age : ℕ  
   parent : Maybe User  
   grantsPermission : User → ItemRequest → Bool  
   isParent : User → Bool

record Transport : Set where  
 field  
   name : String  
   protocol : Protocol

record Context : Set where  
 field  
   name : String  
   timeOfDay : ℕ

record Service : Set where  
 field  
   name : String  
   isApproved : Bool
\end{lstlisting}
We also provide an equivalence operation over Users to be used when looking for parents.  
\begin{lstlisting}
data _≡U_ : Rel User (# 0) where  
 *≡U* : ∀ ( p q : User ) → (User.name p) ≡ (User.name q) → p ≡U q

_≟U_ : (a : User) → (b : User) → Dec ( a ≡U b)  
a ≟U b with (User.name a) Data.String.Properties.≟ (User.name b)  
...       | yes y = yes (*≡U* a b y)  
...       | no  n = no (nent n)  
 where  
   nent : ∀ {s t : User} → (User.name s) ≢ (User.name t) 
                         → ¬ (s ≡U t)  
   nent neq (*≡U* s t ff) = ⊥-elim (neq ff)  
\end{lstlisting}
This assumes names are unique. In a realistic implementation we expect these to explicitly be unique identifiers or public keys.

Next we provide policies as propositional dependent types. These are the equivalent of Rego’s rules. 
\begin{lstlisting} 
data SafeContext : Context → Set where  
 *contextProp* : (c : Context) 
                → (Context.timeOfDay c) Data.Nat.Base.≤ 12   
                → SafeContext c

data HasPermission : (s : User) → (p : ItemRequest) → Set where  
 *hp* : (s : User) → (p : ItemRequest) → (r : User)  
       → (User.isParent s r) ≡ Bool.true   
       → ((User.grantsPermission r) s p) ≡ Bool.true  
       → HasPermission s p

data SafeSender : User → Set where  
 *senderPropOldEnough* : (s : User) → (p : ItemRequest)  
        → (ItemRequest.ageLimit p) Data.Nat.Base.≤ (User.age s) 
        → SafeSender s  
 *senderPropYoung* : (s : User) → (p : ItemRequest) 
        → HasPermission s p → SafeSender s

data SafeChannel : Transport → Set where  
 *isHttps* : (c : Transport) → (Transport.protocol c) ≡ https 
                             → SafeChannel c

data SafePayload : ItemRequest → Set where  
 *safePayload* : (i : ItemRequest) → SafePayload i

data SafeService : Service → Set where  
 *safeService* : (s : Service) → (Service.isApproved s) ≡ Bool.true 
                               → SafeService s

data SafeResponse : Item → Set where  
 *safeResponse* : (r : ItemRequest) → (i : Item)   
     → (ItemRequest.videoName r) ≡ (Item.name i)  
     → SafeResponse i
\end{lstlisting}
The most interesting of these is \texttt{SafeSender}.  There are two rules, following the two allowed conditions.  First is \texttt{*senderPropOldEnough*} which holds if a \texttt{User}’s age is greater or equal to the \texttt{ageLimit} of the requested \texttt{Item}.  Otherwise we have \texttt{*senderPropYoung*} where \texttt{SafeSender} holds if the sender \texttt{HasPermission} to request the \texttt{Item}.  \texttt{HasPermission} holds for a sender and a requested \texttt{Item} if there is another \texttt{User}, that \texttt{User} is the parent of the sender, and that parent grants permission for the sender to see the \texttt{Item}.

The core of this is \texttt{safeCall}, a function that only takes dependent pairs and performs the actual server call underneath.  
\begin{lstlisting}
safeCall : Σ[ c ∈ Context ] (SafeContext c) →  
          Σ[ u ∈ User ] (SafeSender u) →  
          Σ[ ch ∈ Transport ] (SafeChannel ch) →  
          Σ[ s ∈ ItemRequest ] (SafePayload s) →  
          Σ[ s ∈ Service ] (SafeService s) →  
          Maybe Item  
safeCall context sender channel payload service = answer  
where  
   doCall : Transport → ItemRequest → Service → Maybe Item  
   doCall a b c = -- The server call happens here  
   answer : Maybe Item  
   answer with doCall (proj₁ channel) (proj₁ payload) 
                                      (proj₁ service)  
   ...     | nothing = nothing  
   ...     | just result with constrainResponse response 
                                        (proj₁ payload)  
   ...                     | nothing = nothing  
   ...                     | just isGood = just (proj₁ isGood)  
\end{lstlisting}
Each parameter to \texttt{safeCall} is a dependent pair proving the first object has passed the policies applied to it.  Since the actual server call is embedded in \texttt{safeCall}, the call doesn’t occur unless the policy is met.

Accessing \texttt{safeCall} is \texttt{preCall} which just tries to gather those proofs.  
\begin{lstlisting}
preCall : Context → User → Transport → ItemRequest → Service 
                                    → Maybe Item  
preCall context sender channel payload service 
        with (checkContext context)  
...  | nothing = nothing  
...  | just safeContext with (checkSender sender payload)  
...    | nothing = nothing  
...    | just safeSender with (checkChannel channel)  
...       | nothing = nothing  
...       | just safeChannel with (constrainPayload payload)  
...           | nothing = nothing  
...           | just safePayload with (checkService service)  
...              | nothing = nothing  
...              | just safeService = safeCall   
                       safeContext safeSender    
                       safeChannel safePayload safeService  
\end{lstlisting}
It is a cascade of calls to check each proof, which (following intuitionistic logic) either returns \texttt{just} a proof or returns \texttt{nothing}. If all the proofs are available it can call \texttt{safeCall} with the appropriate dependent pairs. One could also replace \texttt{Maybe} with a type containing just a value or \texttt{error} and a message

In between are a number of boring routines to generate the necessary proofs, such as  
\begin{lstlisting}
checkChannel : (s : Transport) 
                → Maybe (Σ[ s ∈ Transport ] (SafeChannel s))  
checkChannel s with (Transport.protocol s) ≟P https  
...             | yes p = just (s , *isHttps* s p)  
...             | no _ = nothing   
\end{lstlisting}
Eventually these should be auto generated.

Given a small set of objects:
\begin{lstlisting}
payload : ItemRequest  
payload = record { name = "Payload" ; 
                   videoName = "I'm PG 13"; ageLimit = 13 }

response = record { name = "I'm PG 13" ; ageLimit = 13 }

daddyPerm : User → ItemRequest → Bool  
daddyPerm child item with (User.name child) ≟S "Sender"  
...          | no _ = Bool.false  
...          | y with (ItemRequest.videoName item) ≟S "I'm PG 13"  
...                | no _ = Bool.false  
...                | yes p = Bool.true  
daddy : User  
daddy = record { name = "Daddy" ; age = 40 ; parent = nothing ;  
                 grantsPermission = daddyPerm ; 
                    isParent = λ { _ → Bool.false }  
 }

isParently : (a : User) → (b : User) → Bool  
isParently a b with a ≟U b  
...             | no _ = Bool.false  
...             | yes _ = Bool.true

sender = record { name = "Sender" ; age = 10 ; 
                  parent = just daddy ;  
                 grantsPermission = λ { u i → Bool.false }  
                 ; isParent = λ { pq → isParently pq daddy }  
 }

youngSender = record { name = "YoungSender" ; age = 10 
                       ; parent = just daddy 
                       ; grantsPermission = λ { u i → Bool.false }  
                       ; isParent = λ { pq → isParently pq daddy }  
 }

channel : Transport  
channel = record {name = "Channel" ; protocol = https }

context : Context  
context = record { name = "Context" ; timeOfDay = 10 }

service : Service  
service = record { name = "Foo" ; isApproved = Bool.true }

service2 : Service  
service2 = record { name = "Foo" ; isApproved = Bool.false }  
\end{lstlisting}
We can evaluate a few calls.  For example, \texttt{preCall context sender channel payload service} will evaluate to \texttt{just response} (while \texttt{sender} is underage, they have permission from \texttt{daddy}), but \texttt{preCall context youngSender channel payload service} fails, as \texttt{youngSender} does not have permission.  Likewise, using \texttt{service2} will fail as it is not approved.

This simple example provides a framework for more sophisticated examples. Like Rego, this entire system can be run as a sidecar in conjunction with the basic application. A client communicates values for \texttt{sender}, \texttt{receiver}, etc., to a script. The script finds the appropriate attributes and evaluates them. Unlike Rego, however, the script could perform the actual service call.  This allows the script to operate on both the payload and response. For example, the payload may contain information requiring a higher security level to view than that of the service; the script can filter out, mask, or tokenize values in the payload to lower its security level. Likewise, the return value may need to be manipulated before being returned to the caller.

Agda being a functional language, we have a main routine in the IO Monad pulling invocations from a \texttt{Stream}. An invocation consists of identifiers for \texttt{sender}, \texttt{receiver}, etc., as well as some representation of the payload. Within the IO Monad we can retrieve or generate attributes for all of these, evaluate the policy functionally, and then perform the call within the IO Monad.

\subsection{Building a decentralized authorization ecosystem}

In a distributed system, if a call crosses a trust boundary, the recipient must ensure any policies are applied, even if the sender has already proven the receiver’s policies were upheld.  
However, if a policy is viewed, as we’ve shown, as the validation of a proposition over some set of properties, and the sender can send some trustworthy values for those properties, which we will now call claims, along with a sketch of the proof, the receiver can just validate the correctness of the proof. 

Some claims, such as claims about the payload itself, may be easy to validate, but others may be claims supplied by third parties.  For example, a claim containing a purchase limit, credit score, or current age attribute is a claim by some third party, such as a bank.  For the sender to pass this to the receiver, the claim needs to be authentic.  In the current web architecture, this implies either a record on a blockchain or a token signed with the private key of the third party.  

Consider the work of the W3C Verifiable Claims WG\cite{vcwg} on their data model\cite{vcmodel}. In the model, 
a claim is a statement about a subject, expressed as a \emph{subject-property-value} relationship.  A credential is a set of claims about the same subject, 
and a \emph{verified} credential is a credential packaged with some metadata and cryptographic proof of the issuer.  Finally, a presentation takes this 
one step further, packaging a set of verified credentials from a single party (the “holder”) with, again, cryptographic proof they come from the holder.

For our current purposes, we will just consider a credential packaging a single claim about an entity or the context. We will posit the existence of a set of \texttt{ClaimServers} at particular URLs. A qualified entity can query a \texttt{ClaimServer} to get information about an entity. Every party to a transaction will have a list of \texttt{ClaimsServer}s they trust, and so would trust a verified claim signed by that \texttt{ClaimsServer}.

Clearly there is a separate set of interactions around becoming qualified to query a \texttt{ClaimServer} and further complexity around which claims for which entities a specific \texttt{ClaimServer} can make.

For two parties to do business, there must be some overlap in the ClaimServers they trust and further overlap in which claims for which entities they trust a particular \texttt{ClaimServer}. For our purposes we will elide that code and assume there’s a single \texttt{ClaimServer} all parties trust.

We add new types supporting trust of the form  
\begin{lstlisting}
data ITrustYou : (s : Service) → (a : ClaimServer) → Set where  
  iTrustYou : (s : Service) → (a : ClaimServer) → ITrustYou s a  
\end{lstlisting}
showing an entity, in this case a \texttt{Service}, trusts some \texttt{ClaimServer}.

A verified claim is a subject-property-value triple signed by a \texttt{ClaimServer}. Within Agda it is a dependent pair.  The first part of the pair is the signed claim, a chunk of text. The second part is a type with three fields. The name of the type reflects the property the claim is about, and the fields are the \texttt{ClaimServer}, the entity, and the value.

For example, when a \texttt{ClaimServer} is queried about a \texttt{User}’s age, this propositional type  
\begin{lstlisting}
data AgeClaim : ClaimServer → User → Set where  
  ageClaim : (a : ClaimServer) → (e : User) → (n : ℕ) 
             → AgeClaim a e  
\end{lstlisting}
represents an \texttt{AgeClaim} and says the \texttt{ClaimServer} says the \texttt{User} is \emph{n} years old. All our properties of interest can be specified as claims. The signed claim is kept for further communication, but the instantiated object can be used in proof.

Proofs are now longer as they need to specify the trust relationships. For example, the proof that a sender was old enough to view an item was  
\begin{lstlisting}
 *senderPropOldEnough* : (s : User) → (p : ItemRequest)  
               → (ItemRequest.ageLimit p) Data.Nat.Base.≤ 
                         (User.age s)   
               → SafeSender s  
\end{lstlisting}
But now it becomes  
\begin{lstlisting}
 *senderPropOldEnough* : (u : User) → (a : ClaimServer) 
                    → (s : Service) → ITrustYou s a 
                    → (ac : AgeClaim a u) → (p : ItemRequest)  
                    → (ItemRequest.ageLimit p) Data.Nat.Base.≤ 
                                   (getAge ac)   
                    → SafeSender s  
\end{lstlisting}
i.e., we have a \texttt{User}, a \texttt{ClaimServer}, and a \texttt{Service}, 
the \texttt{Service} trusts the \texttt{ClaimServer}, the \texttt{ClaimServer} 
makes a claim about the \texttt{User}’s age (\texttt{ac}), and the age specified by the claim is greater or equal to the age specified by the \texttt{ItemRequest}.

Any party can represent its view of a transaction by an Agda record type.  The record has fields for all the appropriate parties and fields for each of the proofs that party requires. For example,   
\begin{lstlisting}
record ServerSide : Set where  
 field  
   callContext : Context  
   callSender : User  
   callTransport : Transport  
   requestedItem : ItemRequest  
   recipient : Service  

   safeContext : SafeContext callContext  
   safeSender : SafeSender callSender  
   safeChannel : SafeChannel callTransport  
   safePayload : SafePayload requestedItem  
   safeService : SafeService recipient  
\end{lstlisting}
The \texttt{ServerSide} record type has fields for the entities of interest in the first part, as we’ve discussed so far. The bottom half has fields with propositional types referring to fields defined in the record.  To instantiate a \texttt{ServerSide} record, one not only needs entities of the appropriate types, but one needs to also provide the necessary proofs about those entities. A \texttt{ClientSide} record would include a field for the returning \texttt{Item} and at least one accompanying proof. 

Other definitions, such as data type definitions, can be included in shared Agda modules.

Now clients can generate appropriate records and send them to the server. The work of validating a policy is shifted from server to client. However, we have no simple serialization mechanism to implement this. Instead we need to convert this for transport ourselves. 

For the entities in the first part of the record, each should have a unique URI. The pair (field name, URI) should be sufficient for the recipient to create a local object of the appropriate type (as the record type gives the type of each field). For the proofs we hope to rely on Agda’s internal reflection mechanism, which provides more than enough information to glean the proof tree. Of course, not all information is necessary. We need the tree of proof constructors and the claims used. In the process of serializing, some information local to where the proof is being executed may need to be signed to verify to the recipient that it was generated by the sender. If reflection turns out to be inadequate, we can alter the code that finds the proof to add in that serialization, so in the end we either have a nothing or a proof with its serialization and the list of verified claims used in generating it.

With this information the recipient will be able to reconstruct a proof from the proof tree and the given claims without needing to either find the proof or perform the process of gathering the claims itself, as the claims are all signed by trusted \texttt{ClaimServers}.

This provides a framework for a decentralized trust architecture, but also a means for collaborating parties to share policy definitions and allow each party to specify their requirements.

\subsection{Beyond Synchrony}

We have described a simple synchronous call and response architecture and a decentralized authorization system. The latter, as well as many financial systems, require a more complex architecture, capable of juggling multiple overlapping communications. We will sketch such an architecture.

Consider a credit card purchase request. Nominally, this has five parties involved - the purchaser, the merchant, the merchant’s bank, the card issuer, and the card network. The dance starts when a merchant sends purchase details to its bank. The merchant’s bank sends these details to the card network, which sends them to the card issuer. If all is above board, the purchase details were precipitated by an exchange between the purchaser and the merchant, but perhaps there is fraud. What should the card issuer do?

Clearly the issuer’s policy will require answering many questions, such as:

- What is the issuer’s history with the merchant? Have they ever been seen? Any past examples of fraud?  
- What is the size of the purchase?  
- Was it point-of-sale or ordered over the internet?  
- Where is the card holder? Can we contact them?  
- … and so on.

Answering these questions, vital to the business, may cause some delay, during which other requests may come in from the merchant, the purchaser, or even the merchant’s bank.  Should these be allowed to proceed, or be paused? If paused, how do we communicate that across requests?

\subsection{A dependently typed response}

In the dependently typed functional world we abandon our synchronous architecture. Instead, as has really been the case, we adopt message passing. Messages come in from an unbounded Stream and are handled quickly.  Perhaps more messages are sent to be later collected before sending a final approval or denial message back to the original sender, after using the collected evidence to evaluate approval or denial against the issuer’s policies.

In the functional world these messages and their processing are handled by a State Monad which is continuously updated before the next message is seen. Some messages, such as a new purchase request, start a new process, but others, such as a response from another system (such as credit scoring), may just update the continuation of an existing process. Sending a message to retrieve a value updates the State with a new continuation to handle the response. Eventually all information is gathered, a response generated, and that process disappears from the State.

When a request comes in requiring additional information, the State can be updated to prevent other transactions from moving forward before that information is gathered and resolved. A policy may place a sliding window transaction limit on an account; the sum of transactions for that window becomes an important attribute. A transaction may pass through a number of states as information is gathered and evaluated; these states can be embedded directly into the types to prevent erroneous messages.

This kind of event loop programming is very common in user interfaces where a user may update any visible item. Techniques such as those in [my dissertation] can be used to lower the burden.

Underlying this approach needs to be a persistent store, whether database, blockchain, or persistent queue. The State Monad retains information between messages; if there is a failure that information would be lost. Providing a persistent store can also allow different executing policy scripts to share information about parties in different transactions.

The format of this store can be derived from the data of the attributes as understood by the scripting language, e.g. Agda. The policies determine which attributes are needed to prove or disprove a policy, and which attributes need to persist across messages. 

In a complex system such as credit card processing, the evidence we’ll need to prove policies will come from a variety of systems, such as bank records, internal credit scoring systems, balances, etc.  Nevertheless, each logical statement in a policy specifies the kind of evidence it needs. The code in the script must have access to routines supplying that evidence or a script will not type check. Any pass through a script leading to a claim a policy is applied needs to find all the necessary evidence.

The discussion above leads directly to a particular approach to policy. Policies are statements of conditions; under certain conditions, certain operations are permitted. As such they are expressible as logical statements. In Agda (and other dependently typed languages serving as theorem provers and programming languages) we can express these policies in a logical language. We can prove properties of a policy directly using Agda as a proof assistant. Policies require kinds of evidence; the policies in Agda specify exactly what kind of evidence is needed. From there one can implement functions to retrieve that evidence. Finally code can be written or likely generated to gather evidence and prove a given operation is valid or invalid under the appropriate conditions. All this together provides very strong guarantees of correctness.

We can go beyond this. As multiple parties are involved in long running financial interactions, should we not consider implementing digital contracts this way?

\subsection{A Rego response}

Rego performs as an oracle with respect to its clients. In theory, using Rego splits policy from processing; code doesn’t need to know about policy, just needs to get an answer. However, when a client queries Rego it needs to ensure Rego has all the information it needs. As described in the Rego documentation, as that becomes more complex, the processes surrounding Rego to ensure its knowledge base is up to date becomes more baroque. As we’ve seen, proving is policy is being correctly applied requires various pieces of evidence. In the dependently typed setting, it is obvious what that evidence is and the script can retrieve evidence as necessary. For Rego, all that information needs to be supplied before at query time, meaning that knowledge needs to be exported from the script and into the supporting infrastructure so all possible evidence is available. 

We have shown how gathering evidence and applying policies can become quite complex. By functioning only as an oracle, Rego can only function in such a scenario by deep cooperation with supporting code. For example, when approving a transaction is based on the status of other outstanding transactions, the code supporting Rego needs to understand how to provide and update that kind of information, leading to further integration. As Rego does not control the actual communication, it cannot operate on the response before the client. The client needs to understand the various stages of a complex communication and access Rego multiple times as the state of the process evolves. 

Inevitably, to keep policy matters away from application code, there will need to be another layer of code that sits between the application and Rego, accessing all the necessary code to add data to the final Rego call and perhaps performing some updates based on the results before returning the final decision to the application. All of this coordination is based on documents flowing between developers of these three components, little of which can be type checked and none of which can be proven correct.

\subsection{Other programming languages}

What we have demonstrated is based on the particular characteristics of dependently typed languages; it would not generally hold for other languages as it would require a change to the type system. A notable exception is Java’s allowance for type annotations and its integration with the Checker Framework\cite{checkerframework}.

Java has included annotations since version 1.5 and extended those to types in SE 8. Annotations on types allow developers to add additional constraints on types. At the same time, Java added a plugin framework for the compiler. The Checker Framework (CF) is a compiler plugin enabling users to extend compiler functionality by processing typed annotations at compile time.

CF ships with many annotations, such as a nullness checker, a lock checker, and, closest to this work, a tainting checker. Tainted and untainted values can be seen as private and public, controlling visibility.

With this assist, annotations on Java code can be used to force various operations to be performed with compile time checks, including with relationships among annotations. It can also be used to ensure callouts to policy code, such as Agda or other, and that the results are properly handled. Otherwise a policy framework is entirely dependent on the client to actually implement the decisions of the policy server.

Python has powerful decorators to provide some of the services of annotations, although only at run time. Nevertheless 
these have been used to provide metadata to numerous services, such as the Pydantic\cite{pydantic}
type checker to support sophisticated but optional type hints. It would be interesting to combine that with our proposals.

\section{Conclusion}

Current approaches to the access control problem rely on ad hoc policies written in untyped languages. As the policy is implemented in the body of rules, there is no way to test correctness except through observation and testing use cases. There is also no way to separate the specification of the policy and the implementation code checking individual requests.

We demonstrate the applicability of dependent types, especially propositional types, to the access control problem. By specifying policies directly as types in a strongly typed programming language we get the usual compile time guarantees that our policies are well formed as types. However, a policy is also a set of requirements to be fulfilled before an action is taken. Our approach ensures these steps are taken through compile time guarantees; i.e., specifying the policies as types forces the code validating an action to necessarily perform all the necessary steps or it will not compile; the code needs to always prove the policy has been enforced or it does not type check.

Having specified a policy as a set of propositions separate from its implementation, we can directly analyze the policy.  Agda, like other dependently typed languages, such as Lean, doubles as a proof assistant. We can use this to prove desirable properties of our policies, such as preventing simultaneous withdrawals from an account. By including these proofs in our code, we can test for regressions when policies evolve. A regression proof provides much stronger guarantees than any number of regressions tests, the only way to test existing systems. We can also translate the policy into other formats so we can apply both SAT solvers and model checking to provide guarantees.

As there is no separate specification of a policy in Rego, there is no separate specification of the attributes required to validate that policy. Therefore there will be a need for an intermediate layer between application and script to provide them. In Agda, attributes are well defined and with an API the script can retrieve the values it needs.

As policies become more complex, there are interactions among transactions due to race conditions, account limits, time delays, etc. These require maintaining state across invocations. We show how to apply functional programming techniques to minimize the overhead of checking these dependencies, as well as interacting with external transactional stores to share state. We believe the intermediate layer mentioned will become more complex with existing policy languages, breaking down the application/policy barrier.

Currently, any service receiving a message needs to apply policies from a no trust perspective. However we show how to integrate our approach with a verifiable claims infrastructure, such as that specified by the W3C Verifiable Claims WG\cite{vcwg}. In such a system, the client can undertake the overhead of retrieving attributes and proving claims. The proof sketch and signed, verified attribute values can be passed to the server, which only needs to verify the proof and not start from the beginning.

This implies the ability to share definitions of attributes and policies among the entities involved in some transaction, such as a digital contract. We show how Agda language constructs, such as modules and record types, can serve to specify these, as well as share proofs and back and forth.  

\bibliography{refs}

\begin{thebibliography}{17}
\providecommand{\natexlab}[1]{#1}
\providecommand{\url}[1]{\texttt{#1}}
\expandafter\ifx\csname urlstyle\endcsname\relax
  \providecommand{\doi}[1]{doi: #1}\else
  \providecommand{\doi}{doi: \begingroup \urlstyle{rm}\Url}\fi

\bibitem[alf()]{alfa}
{ALFA YAML Profile for XACML}.
\newblock \url{https://alfa.guide}.

\bibitem[che()]{checkerframework}
{The Checker Framework}.
\newblock \url{https://checkerframework.org/}.

\bibitem[opa()]{opa}
{Open Policy Agent — Rego Documentation}.
\newblock \url{https://www.openpolicyagent.org/docs/policy-language}.

\bibitem[ope()]{openfga}
{OpenFGA — Fine-grained Authorization}.
\newblock \url{https://openfga.dev/}.

\bibitem[ory()]{ory}
{Ory – Open Source Identity Infrastructure}.
\newblock \url{https://www.ory.sh/}.

\bibitem[pyd()]{pydantic}
{Pydantic — Data Validation and Settings Management}.
\newblock \url{https://docs.pydantic.dev/latest/}.

\bibitem[sen()]{sentinel}
{HashiCorp Sentinel}.
\newblock \url{https://www.hashicorp.com/sentinel}.

\bibitem[sty()]{styra}
{Styra Inc.}
\newblock \url{https://www.styra.com/}.

\bibitem[vcw()]{vcwg}
{W3C Verifiable Credentials Working Group}.
\newblock \url{https://www.w3.org/2017/vc/}.

\bibitem[Cutler et~al.(2024)Cutler, Disselkoen, Eline, He, Headley, Hicks,
  Hietala, Ioannidis, Kastner, Mamat, McAdams, McCutchen, Rungta, Torlak, and
  Wells]{cedar}
Joseph~W. Cutler, Craig Disselkoen, Aaron Eline, Shaobo He, Kyle Headley,
  Michael Hicks, Kesha Hietala, Eleftherios Ioannidis, John Kastner, Anwar
  Mamat, Darin McAdams, Matt McCutchen, Neha Rungta, Emina Torlak, and
  Andrew~M. Wells.
\newblock Cedar: A new language for expressive, fast, safe, and analyzable
  authorization.
\newblock 8\penalty0 (OOPSLA1), 2024.
\newblock URL \url{https://doi.org/10.1145/3649835}.

\bibitem[de~Moura et~al.(2024)de~Moura, Ullrich, et~al.]{lean4book}
Leonardo de~Moura, Sebastian Ullrich, et~al.
\newblock \emph{Theorem Proving in Lean 4}.
\newblock 2024.
\newblock Accessed 31~May~2025.

\bibitem[Pang et~al.(2019)Pang, Caceres, Burrows, Chen, Dave, Germer, Golynski,
  Graney, Kang, Kissner, Korn, Parmar, Richards, and Wang]{zanzibar}
Ruoming Pang, Ramon Caceres, Mike Burrows, Zhifeng Chen, Pratik Dave, Nathan
  Germer, Alexander Golynski, Kevin Graney, Nina Kang, Lea Kissner, Jeffrey~L.
  Korn, Abhishek Parmar, Christina~D. Richards, and Mengzhi Wang.
\newblock Zanzibar: Google’s consistent, global authorization system.
\newblock In \emph{2019 {USENIX} Annual Technical Conference ({USENIX} {ATC}
  '19)}, Renton, WA, 2019.
\newblock URL
  \url{https://research.google/pubs/zanzibar-googles-consistent-global-authorization-system/}.

\bibitem[Rij\-ke(2022)]{rijke-hott}
Egbert Rij\-ke.
\newblock \emph{Introduction to Homotopy Type Theory}.
\newblock 2022.

\bibitem[Sporny et~al.(May 2025)Sporny, Longley, Chaddwick, and
  Herman]{vcmodel}
Manu Sporny, Dave Longley, David Chaddwick, and Ivan Herman.
\newblock {Verifiable Credentials Data Model v2.1}.
\newblock May 2025.

\bibitem[TC(January 2013)]{xacml}
OASIS~XACML TC.
\newblock {eXtensible Access Control Markup Language (XACML) 3.0}.
\newblock January 2013.

\bibitem[Team(2025)]{idris2tutorial}
The Idris 2~Development Team.
\newblock \emph{Idris 2 Tutorial}.
\newblock 2025.
\newblock Accessed 31~May~2025.

\bibitem[Wadler and Kokke(2023)]{plfa}
Philip Wadler and Wen Kokke.
\newblock \emph{Programming Language Foundations in Agda}.
\newblock 2023.
\newblock Accessed 31~May~2025.

\end{thebibliography}

\end{document}